\documentclass{IEEEtran}

\usepackage[latin1]{inputenc}
\usepackage[english]{babel}
\usepackage{times}
\usepackage{graphicx}

\def\thealg{\textit{\textbf{PROSA}}}

\title{Social behaviours applied to P2P Systems: An efficient algorithm for 
	resources organisation} 
\author{
	Vincenza Carchiolo\hspace{5mm} Michele Malgeri\hspace{5mm} 
	Giuseppe Mangioni\hspace{5mm} Vincenzo Nicosia\\
	\em{Dipartimento di Ingegneria Informatica e delle Telecomunicazioni}\\ 
    \em{Facolt\`a di Ingegneria - Universit\`a di Catania} \\
    \em{V.le A. Doria, 6 - 95125 Catania (ITALY)}\\
	{\{car, mmalgeri, gmangioni, vnicosia\}@diit.unict.it}
}

\begin{document}

\pagestyle{empty}

\maketitle
	
	\begin{abstract}
	

P2P systems are a great solution to the problem of distributing
resources. The main issue of P2P networks is that searching and
retrieving  resources shared by peers is usually expensive and does
not take into account similarities among peers. In this paper we present
preliminary simulations of \thealg, a novel algorithm for P2P network structuring,
inspired by social behaviours. Peers in \thealg{} self--organise in
social groups of similar peers, called ``semantic--groups'', depending
on the resources they are sharing. Such a network smoothly evolves to
a  small--world graph, where queries for resources are efficiently 
and effectively routed. 

	\end{abstract}
		
	\section{Introduction}

	In the last years social communities have been deeply
	studied not only by psychologists or sociologists, but also by	
	computer scientists. The main point is that social communities seems
	to naturally possess really interesting characteristics that can be
	exploited in computer science. 
	Studying collaboration communities, researchers have found an
	interesting structure that seems to arise whenever a network of
	relationships among entities is involved: the so called
	``small--world'' graphs. A small--world graph is a graph which
	present a high clustering coefficient (i.e. similar peers usually
	link each other) and a relative small average path length (i.e. the
	average number of intermediates between two peers is small).
	
	The small--world property seems to be a characteristic of many
	human communities, such as mathematicians, actors, 	scientists. 
	A small--world arises almost naturally whenever 
	social contacts among people are involved: many researchers are
	trying to understand the reasons of this behaviour.
	In this work we're not interested in answering this question. 
	Our target is just to develop a P2P system using rules and concepts
	inspired by human behaviours and relationships dynamics.

	In a social network there are several kind of links among
	people, from  simple acquaintance to friendship.  Note that usually
	social links are not symmetric: all British people know who is the
	prime	minister of UK, but the prime minister himself doesn't
	directly know all of them. 
	We say that a person has an  ``acquaintance--link'' to somebody else
	if he simply knows him. In real life it is really simple to gain
	acquaintance--links to anybody: a person met on the stairs and a
	taxi driver are examples of such links.

	Nevertheless our social life is mainly based upon
	``semantic--links''. A semantic--link is more than a simple
	acquaintance link, since it requires not only to know a person, but
	also to share with him  knowledge, culture, interests,
	job, hobbies or abilities. We have semantic--links to our parents,
	brothers, friends, colleagues, teachers and so on, because we share
	with them our home (parents), our interests (friends), our job
	(colleagues) and so on. Usually  social life is heavily focused
	on relationships with our ``semantic--links'', since we 
	spend most of our time talking, discussing, working, staying with friends,
	parents, colleagues and so on.

	For this reason we use semantic--links in order to solve
	every day problems: we ask our parents for a suggestion,
	we  talk with	friends about shared interests, we ask a colleague for
	help about finding a bug in a program and so on. 

	Starting from these observations of social relationships dynamics, 
	we defined \cite{PROSA@AP2PC06} a P2P structure, named \thealg, 
	in which similar peers build	``semantic--links'' to each other.
	
	\thealg{} uses a self--organising distributed algorithm that
	dynamically	links peers	sharing similar knowledge and resources,
	putting  them into high	clustered and self--structured ``semantic
	groups''.	Searching and retrieving resources in a semantic group is
	really fast and efficient, since peers into a group are strongly 
	connected.
	As real social communities, \thealg{} naturally
	evolve to a small--world network (as reported below), that
	allows  peers to retrieve	resources in a fast and efficient way,
	also  if the requested resources do not belong to the same semantic
	group  of the requesting peer. 
	
	In this	paper we report some preliminary simulation results of
	\thealg. As showed in Section \ref{sect:result}, the
	distributed algorithm used to route query and to manage links among
	peers, allows to obtain a great percentage of
	successfully answered queries and a small average path length
	between peers. Comparisons with a simple flood strategy and with a
	``random--walk'' are also reported.
	
	The paper is organised as follows: Section \ref{sect:related} is a
	short survey about current work in the field of P2P resource
	retrieval; in Section \ref{sect:idea}	we discuss our proposal;
	 in Section\ref{sect:result} we show simulation results and
	finally Section \ref{sect:future} presents a plan for future
	work.  
	
	\section{Related work}
	\label{sect:related}
	
	\thealg{} is not the first attempt to organise a P2P network taking
	into account ``semantic'' proximity of shared resource in order to
	optimise query routing.
	Some recent works (\cite{SETS}\cite{GES})
	proposed to organise a P2P network in semantic groups of
	``similar'' peers, to facilitate resource search and retrieval
	based on semantic queries. In particular in SETS \cite{SETS}
	the network is split in semantic areas by a super--peer which
	also maintains a table of groups centroids; a centroid
	represents the ``topic'' of a given area. The main drawback of
	SETS is the introduction of a network manager, which
	represents a single point of fault. In GES \cite{GES} peers
	maintains two sets of links to other peers: semantic--links
	and random--links. Queries for resources are first forwarded
	to a so--called ``semantic--target'', which is the first peer
	that can answer the query, and then flooded to this peer
	neighbours (the semantic group). 

	\thealg{} is an early attempt to implement a bio--inspired link
	management algorithm into a pure P2P overlay network. 
	We think it is really interesting to study real networks, such as
	social communities, in order to find new and effective algorithms
	for sharing, searching and distributing resources in P2P
	environments.

	\section{\thealg }
	\label{sect:idea}

	\thealg{} is a P2P network based on acquaintance--
  and semantic--links, where peers join the network in a way
  similar to a ``birth'', then achieve more links to other peers
  according to the social model, i.e. by linking (semantically)
  with peers which have similar interests, culture, hobbies,
  works and so on, and maintaining a certain number of
  ``random'' acquaintances. In P2P networks the culture or
  knowledge of a peer is represented by the resources
  (documents) it shares with other peers. On the other hand,
  different types of ``links'' among peers simulate
  acquaintances and semantic--links.  To implement such a model
  it is necessary to have:
	\begin{itemize}
	\item A system to model knowledge, culture, interests etc...
  \item A self--organising network management algorithm
	\end{itemize}

	\subsection{Modelling Knowledge}
	\label{subsect:vsm}
	In \thealg, knowledge (each resource) is modelled through
  Vector Space Model (VSM)~\cite{TFIDF} . In this
  approach each document is represented by a state--vector of
  (stemmed) terms called Document Vector (DV); each term in the
  vector is assigned a weight based on the relevance of the term
  itself inside the document.  This weight is calculated using a
  modified version of TF--IDF \cite{866292} schema, as follows:
	$$
	w_{t,D}= 1 + \log(f_{t})
	$$
	
	where $f_{t}$ is the term frequency into the document.  It has been
	proved \cite{TFIDF} that this way of calculating relevance is a good
	approximation of TF--IDF ranking schema.  The VSM representation of a
	document is necessary to calculate the relevance of a document with
	respect to a certain query.  We model a query by means of a so--called
	Query Vector (QV), that is the VSM representation of the query itself.
	Since both documents and queries are represented by state--vectors, we
	define the relevance of a document (D) with respect to a given query
	(Q) as follows:
	\begin{equation}
		\label{eq:dvqv}
		r(D,Q) = \sum_{t\in D\cap Q} w_{t,D} \cdot w_{t,Q}
	\end{equation}
  
	Using VSM we obtain also a compact description of a
	peer knowledge. This description is called ``Peer-Vector'' (PV), and is
	computed as follows:
	\begin{itemize}
	\item[-]
		For each document hosted by the peer, the frequencies of terms it
		contains are computed (${F_{t,D}}$).
	\item[-]
		Terms frequencies  for different documents are summed together,
		obtaining overall frequency for each term:
		$$
		F_{t} = \sum_{t} F_{t,D}
		$$
	\item[-]
		Then a weight is computed for each term, using:
		$$
		w_{t,P} = 1 + \log(F_t)
		$$
	\item[-]
		Finally all weights are put into a state--vector and the vector is
		normalised.
	\end{itemize}
	The obtained PV is a sort of ``snapshot'' of the peer knowledge,
	since it contains information about the relevant terms of the
	documents it shares.\\

  The relevance of a peer (P) with respect to a given query (Q) is defined as
  follows: 
	$$
	r(P,Q) = \sum_{t\in P\cap Q} w_{t,P} \cdot w_{t,Q}
	$$
  This relevance is used by the \thealg{} query routing algorithm.
	It is worth noting that a high relevance between a QV and a PV
	means that probably the given peer has documents that can match the
	query.
	
	\subsection{Network Management algorithm}
	\label{subsect:alg}
  As stated above, relationships among people are usually based
  on similarities in interests, culture, hobbies, knowledge and
  so on. And usually these kind of links evolve from simple
  ``acquaintance--links'' to what we called ``semantic--links''.
	
	To implement this behaviour three types of links have been introduced:
	\begin{itemize}
	\item[-]
		Acquaintance--Link (AL)
	\item[-]
		Temporary Semantic--Link (TSL)
	\item[-]
		Full Semantic--Link (FSL)
	\end{itemize}
	TSLs represent relationships based on a partial knowledge of a
	peer. They are usually stronger than ALs and weaker than FSLs.
	
\begin{figure}[!htbp]
		\centering
		\includegraphics[scale=0.25]{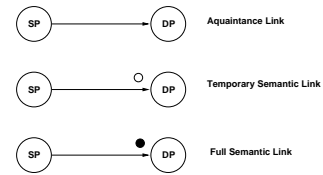}
		\caption{Link types}
		\label{fig:links}
	\end{figure}

	Since usually relationships are not symmetric, it is necessary
  to specify what are the source peer (SP) and destination peer
  (DP) of a link. Figure \ref{fig:links} shows the
  representations for the three different types of links.
	
	Each peer into \thealg{} maintains a list of known peers, that we call
	Peer List (PL). This list contains all the links gained by a peer
	during his ``life''. It is similar to a personal phone book: when we
	meet a person we link to him with an AL. If we share interests,
	knowledge or anything else with him, the AL pointing to him smoothly 
	becomes  a semantic--link. It first evolves to a Temporary Semantic
	Link,  and then to a Fully Semantic Link.

	\subsubsection{Joining}
\begin{figure}
		\centering
		\includegraphics[scale=0.25]{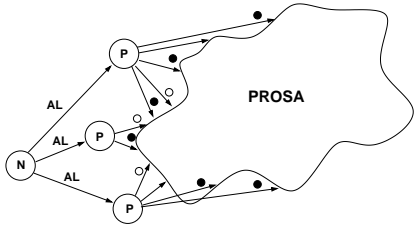}
		\caption{A new node joining \thealg}
		\label{fig:join}
	\end{figure}

	The case of a node that wants to join an existing \thealg{} network is
  similar to the birth of a child. At the beginning of his life
  a child ``knows'' just a couple of people (his parents). A new
  peer which wants to join, just searches other peers (for
  example using broadcasting, or by selecting them from a list
  of peer that are supposed to be up, as in Freenet\cite{clarke01freenet}
	 or Gnutella)
  and adds some of them in his PL as Hals. These are ALs because
  a new peer doesn't know anything about its ``relatives'' until
  he doesn't make query to them for resources. This behaviour is quite
  easy to understand: when a baby comes to life he doesn't know
  anything about his parents. He doesn't know his father's job,
  neither that is mother is a biologist.  The joining phase is
  represented in figure \ref{fig:join}, where ``N'' is the new
  peer; N chose some other peers (P) at random as initial ALs.
	
	\subsubsection{Updating}
	\label{subsect:updating}

	In \thealg{} FSLs dynamics are strictly related to queries. When
  a user of \thealg{} requires a resource, he performs a query and
  specifies a certain number of results he wants to obtain. The
  relevance of the query with respect to the resources hosted by the user's
  peer is first evaluated, using equation \ref{eq:dvqv}. If none
  of the hosted resources has a sufficient relevance with
  respect to the query, the query has to be forwarded to other
  peers. The mechanism is quite simple:
	
	\begin{itemize}
	\item[-]
		A query message containing the QV, a (possible) unique QueryID, the 
		source address and the required number of results is built.
	\item[-]
		If the peer has neither FSLs nor TSL, i.e. it has just AL, the query
		message is forwarded to one link at random.
	\item[-]
		Otherwise, the peer computes the relevance between the query and each 
		entry of his Peers--List. 
	\item[-]
		It selects the link with a higher relevance, if it exists, and
		forwards the query message to it.
	\end{itemize}
	
	When a peer receives a query forwarded by another peer, it first
	updates its PL. If the requesting peer is an unknown peer, a new TSL
	to that peer is added in the PL, and the QV becomes the corresponding
	Temporary Peer Vector (TPV). If the requesting peer is a TSL for the
	peer that receives the query, the corresponding TPV in the list is
	updated, adding the received QV and normalising the result. If the
	requesting peer is a FSL, its PV is in the PL yet, and no updates
	are necessary.
	
	After PL update, the relevance of the query and the peer
  resources is computed.  There are three possible cases:
	\begin{itemize}	
	\item[-] None of the hosted documents has a sufficient relevance. In this case
    the query is forwarded to another peer, using the same
    mechanism used by the forwarder peer. The query message is
    not modified.
	\item[-] The peer has a certain number of relevant documents,
    but they are not enough to full-fill the request. In this
    case a response message is sent to the requester peer,
    specifying the number of matching documents and the
    corresponding relevance. The message query is forwarded to
    all the links in the PL whose relevance with the query is
    higher than a given threshold (semantic flooding). The
    number of matched resources is subtracted from the number of
    total requested documents before forwarding.
	\item[-] The peer has sufficient relevant documents to
    full-fill the request. In this case a result message is sent
    to the requesting peer and the query is no more forwarded.
	\end{itemize}
	
	\begin{figure}[!htb]
		\centering
		\includegraphics[scale=0.25]{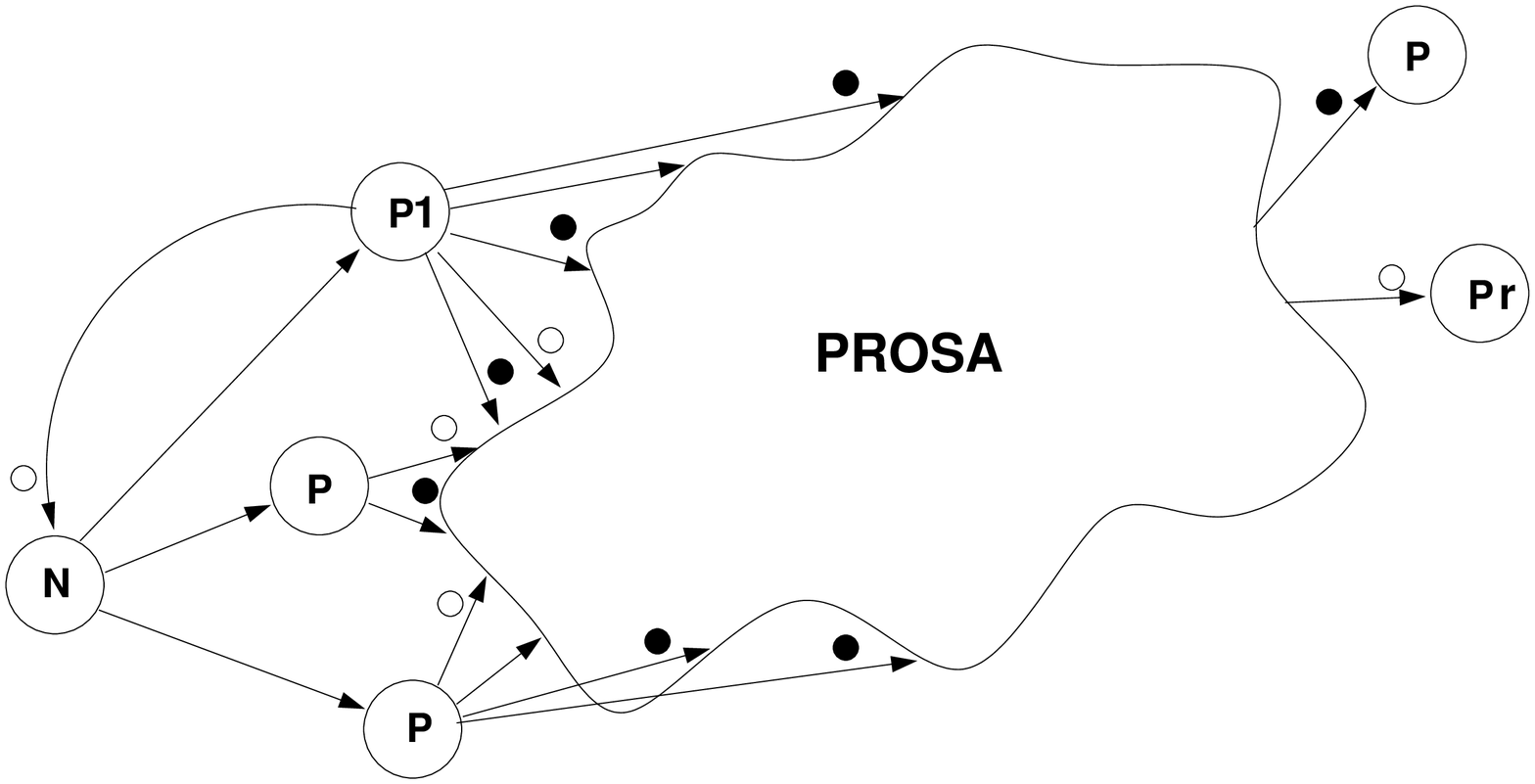}
		\caption{Query forwarding: new TSL arise}
		\label{fig:query_one}
	\end{figure}
	This situation is showed in figure \ref{fig:query_one}, where
  peer ``N'' forwards a query to one of his ALs randomly chosen,
  since it has neither TSLs nor FSLs. In our example the chosen
  peer is ``P1''. As soon as P1 receives the QV, it
  automatically establish a TSL with N (see figure
  \ref{fig:query_one}) and then it forwards the query if needed.
	
\begin{figure}[!htbp]
		\centering
		\includegraphics[scale=0.25]{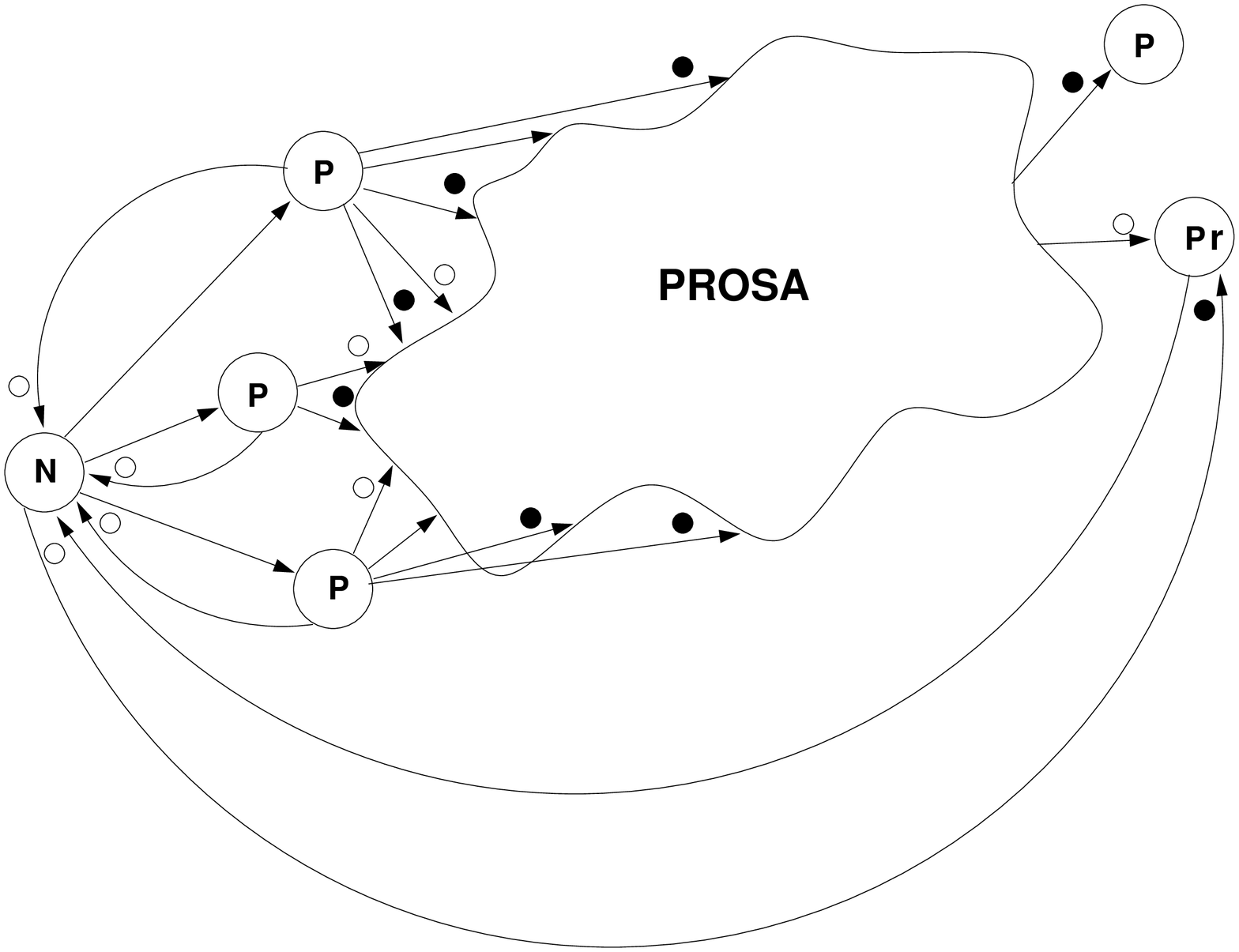}
		\caption{Query forwarding: new FSL arises}
		\label{fig:query_two}
	\end{figure}

	When the requesting peer receives a response message it
  presents the results to the user. If the user decides to
  download a certain resource from another peer, the requesting
  peer contacts the peer owning that resource and asks it for
  download. If download is accepted, the resource is sent to the
  requesting peer, together with the Peer Vector of the serving
  peer. This case is illustrated in figure \ref{fig:query_two},
  where peer ``N'' received a response from peer ``Pr'' and
  decided to download the corresponding resource. Note that Pr
  established a TSL with N, because it received a QV from it,
  and N established a FSL with Pr, because it successfully
  received a resource from it.

	\section{\thealg{} simulations and results}
	\label{sect:result}
	
	The main target of this work is to show that a
	relationships--inspired network naturally evolves to a
	small--world. Simulation results confirm that \thealg{} is a
	small--world network: it presents a high clustering coefficient and
	a small average path length. 

	Since links between peers in \thealg{} are not symmetric, it is possible
	to represent a \thealg{} network as a directed graph G(V,E). The Clustering
	Coefficient for a node ($CC_n$) in a directed graph can be defined as follows:
	
	\begin{equation}
		\label{eq:ccnode}
		CC_{n} = \frac{E_{n,real}}{E_{n,tot}}
	\end{equation}
	
	where $n$'s neighbours are all the peers to which $n$ as linked to,
	$E_{n,real}$  is the number of edges between n's neighbours and
	$E_{n,tot}$ is the maximum number of possible edges between n's
	neighbours. Note
	that if $k$ is in the neighbourhood of $n$, the vice-versa is not
	guaranteed, due to the fact that links are directed.
	The clustering coefficient of a graph ($CC$) is defined as the mean graph
	coefficient for all the vertices (nodes) in the graph:
	\begin{equation}
		CC = \frac{1}{|V|}\sum_{n \in V}{CC_{n}}
	\end{equation}
	
	In figure \ref{fig:prima} the CC and average	path length (APL) of
	\thealg{} is compared to those of the ``equivalent'' random graph
	(\textit{rnd}). Given a graph G(V,E), its equivalent random graph has the
	same number of nodes and edges and a random out-degree distribution.
	
	The $CC$ and the $APL$ of a random graph
	with $|V|$ vertices and $|E|$ edges has been computed using equations
	(\ref{eq:ccrandom}) and (\ref{eq:aplrandom}) \cite{watts98strogatz}.
	
	\begin{equation}
		\label{eq:ccrandom}
		CC_{rnd} = \frac{|E|}{|V| \cdot (|V| - 1)}
	\end{equation}
	
	\begin{equation}
		\label{eq:aplrandom}
		apl_{rnd} = \frac{\log{|V|}}{\log{(|E|/|V|)}}
	\end{equation}
	
	\begin{figure}[!htbp]
		\centering
		\tiny
		\begin{tabular}{|l|l||l|l||l|l||l|}
			\hline
			\textbf{\# nodes} & \textbf{\# edges} &
			\textbf{CC\_prosa} & \textbf{APL\_prosa} & \textbf{CC\_rnd} & 
			\textbf{ALP\_rnd} & \textbf{CC\_prosa/CC\_rnd}\\ \hline \hline
			400 & 15200 & 0.26 & 2.91 & 0.095 & 1.65 & 2.7 \\ \hline
			600 & 14422 & 0.19 & 2.97 & 0.04 & 2.01 & 4.75 \\ \hline
			800 & 14653 & 0.17 & 2.92 & 0.02 & 2.29 & 8.5 \\ \hline
			1000 & 14429 & 0.15 & 2.90 & 0.014 & 2.58 & 10.7\\ \hline
			3000 & 15957 & 0.11 & 2.41 & 0.002 & 4.8 & 55 \\ \hline
			5000 & 19901 & 0.06 & 2.23 & 0.0008 & 6.17 & 75\\ \hline
		\end{tabular}
		\caption{Clustering coefficients and APL for different network size}
		\label{fig:prima}
	\end{figure}

	These measures regard the case of \thealg{} networks where each peer
	starts with 20 documents on average. The $CC$ and $APL$ are computed
	after 10.000 queries. Each query contains 4 terms, on average. A
	query is considered ``successfully'' when at least one matching
	document is found. The maximum number of required document is 5.
	
	Looking at the results, it is clear that \thealg{} networks always present
	a higher clustering coefficient than the corresponding random
	graphs. This means that each peer is linked with a
	strongly connected neighbourhood, which represents (a part of) the
	``semantic group'' joined by the peer. This behaviour is due to
	the fact that  links are mainly ``semantic links'' (both FSLs and TSLs)
	with nodes that provided (or requested) resources belonging to a given
	field. Note also that the APL for a \thealg{} network decreases when the
	number of nodes increases, while it seems to linearly depend on the
	network size for the correspondent random graph. Note that the APL
	for \thealg{} is measures as the average deepness of a query, so it
	represents a very accurate estimation of the real APL.


	\begin{figure}[!htbp]		
	\centering
		\includegraphics[scale=0.25]{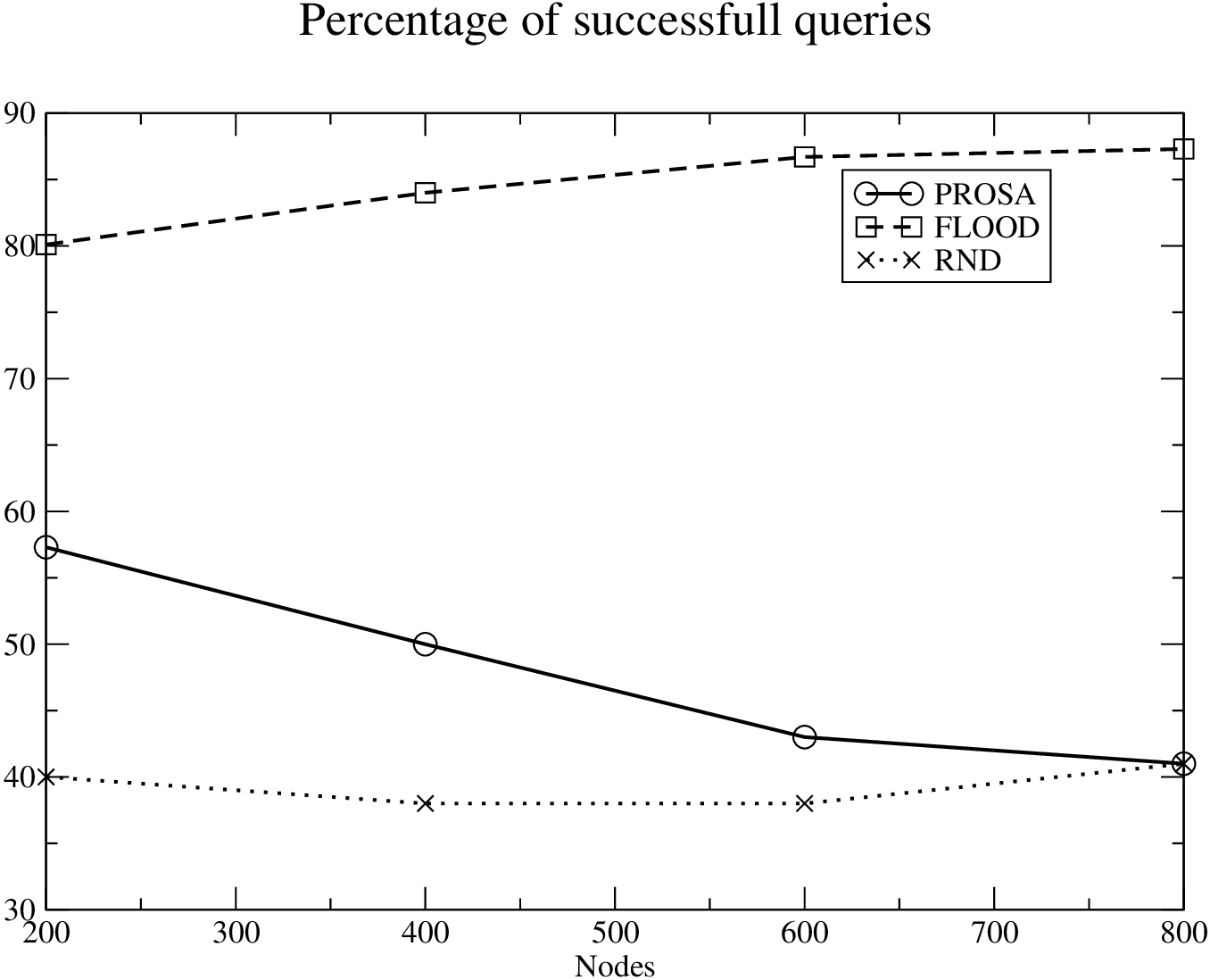}
		\caption{}
		\label{fig:queryok}
	\end{figure}
	
	To evaluate the efficiency of \thealg{} we compared it with a pure 
	flooding network and a random walk network. In a flooding network
	queries are routed using a classical flooding algorithm; in a random
	walk network query are forwarded through randomly chosen links.
	Figure \ref{fig:queryok} shows the rate  of
	successful queries	for \thealg, a pure flooding network and a
	random--walk network.
	The highest percentage is (obviously!)
	obtained with a pure flooding, because in that case the most part of
	the network is visited, and if matching resources do exist, they
	will eventually found.
	\thealg{} results to be more efficient than a random walk for different
	network sizes. This is due mainly to the fact that peers in \thealg{}
	do not forward queries at random, but using the algorithm described in
	Section \ref{subsect:updating}. If a peer does not have relevant
	documents for a given query, it forwards the query to one of the
	peers it links to, choosing the one that has the best ``relevance''
	with the query. This way the query is routed to those
	peers that can (probably) answer it. Note that the efficiency of
	\thealg{} with respect to the percentage of successful queries  is
	related  to the number of query performed by  peers,
	since semantic--links are a side--effect of searching and retrieving
	resources. We obtain a decreasing percentage of successful queries
	when  network size grows, because the total number of queries is the
	same for all simulation reported.

	\begin{figure}[!htbp]
	\centering
	\includegraphics[scale=0.25]{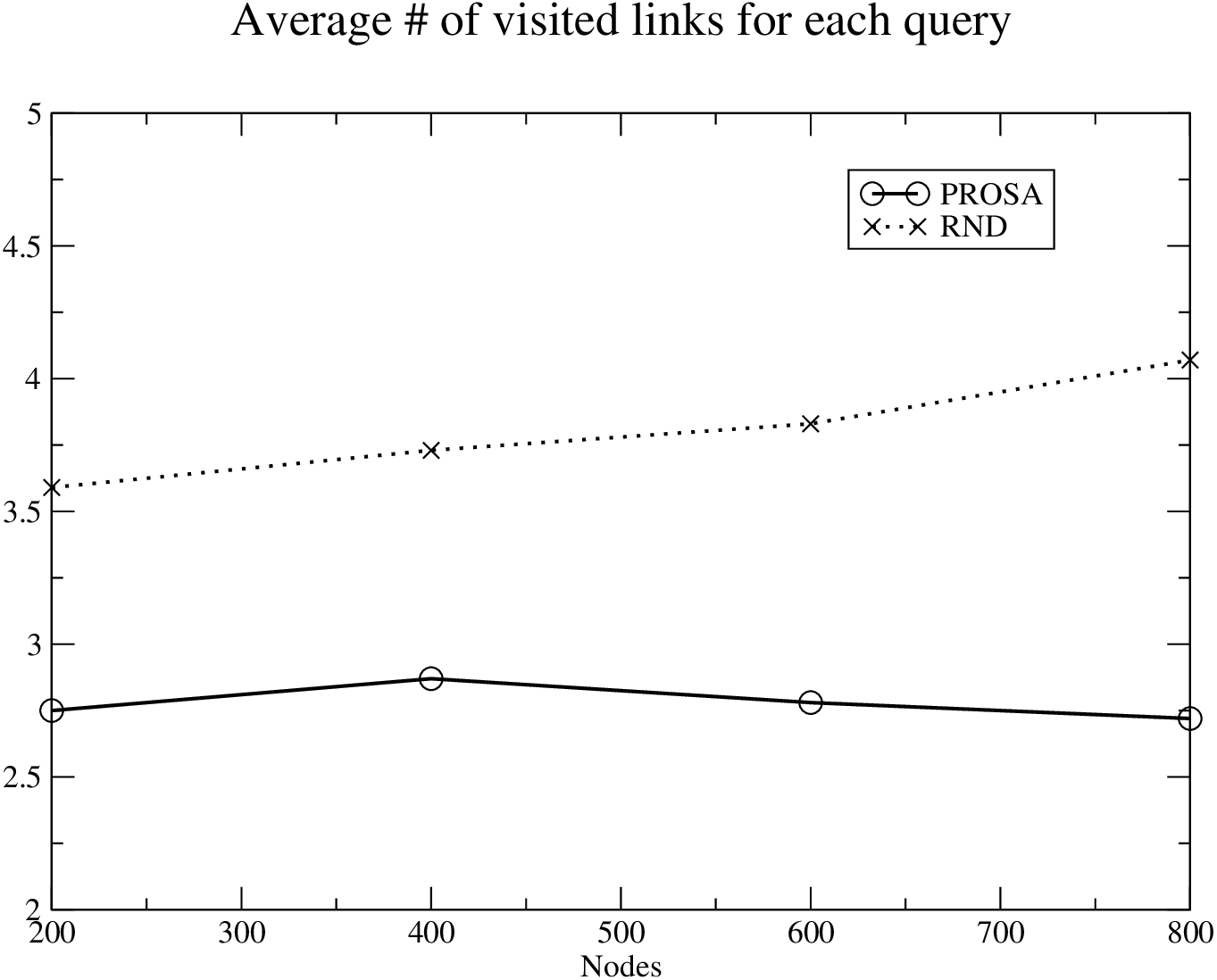}
		\caption{}
		\label{fig:querywalks}
	\end{figure}

	In figure \ref{fig:querywalks} we show the average number of different
	links visited for a successful query, both for \thealg{} and for a
	pure random--walk search.\footnote{Results for pure flooding are
	not reported, since they are from 100 to 150 times larger that those
	of \thealg{} and random walk} Using \thealg{} we obtain a smaller
	average amount  of walks for successful	query than that obtained
	with a pure random search. We can explain this fact as a consequence
	of both the query routing algorithm used by \thealg{} and the link
	updating policy. In \thealg{} new links among similar peers arise
	almost naturally: a new FSL is established for every document gained
	by a peer as a query result. Since a peer interested into a
	particular field usually makes query for resources in that field,
	the higher the number of queries performed, the higher the number of
	new FSL to a specific semantic group. After a (small) amount of
	queries, a peer results to be strongly connected to other peers
	in the same group. New queries will be directly forwarded to the
	best--matching group, in a small number of steps.

	\begin{figure}[!htbp]
		\centering
		\includegraphics[scale=0.25]{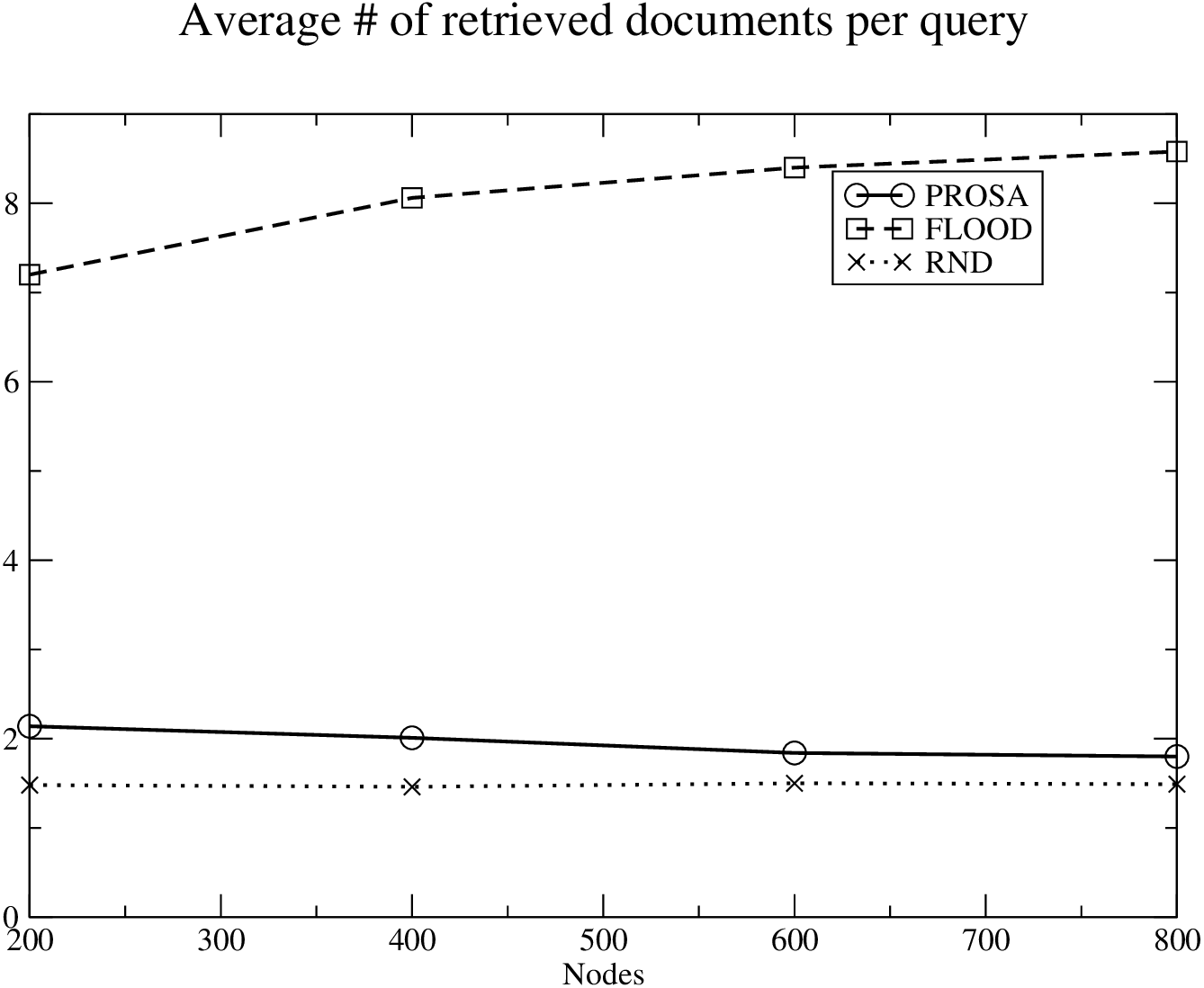}
		\caption{}
		\label{fig:prosafloodrnd}
	\end{figure}
	
	In figure \ref{fig:prosafloodrnd} we show the number of average retrieved
	documents per query for \thealg, a pure flooding--based search
	(such as that implemented in the first version of Gnutella) and a
	simple random--walk. The average number of documents retrieved by
	\thealg{} is always higher than that obtained with a dummy
	random--walk. This is due to the fact that forwarding query at
	random does not guarantees resources to be found. On the other hand,
	if a query is forwarded to a ``relevant'' peer (i.e. a peer that
	contains documents that match the query), it is highly probable to
	obtain a success.
	
	\begin{figure}[!htbp]
		\centering
			\includegraphics[scale=0.25]{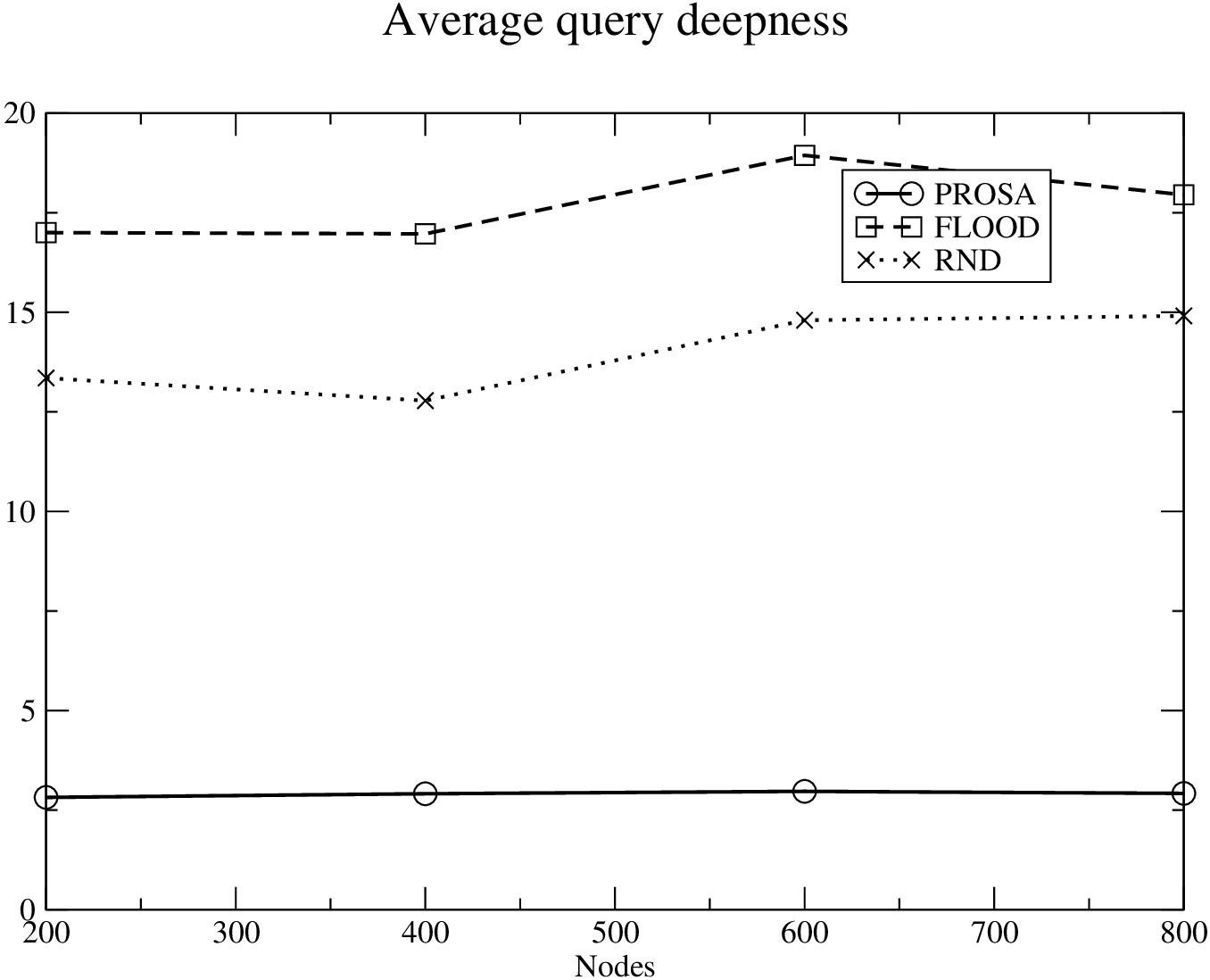}
			\caption{}
			\label{fig:deepness}
	\end{figure}
			
	Finally, figure \ref{fig:deepness} shows the average query deepness,
	i.e. the average number of hops needed to satisfy a query. It is
	clear that the average query deepness for \thealg{} is heavily lower
	than in the case of a pure flooding or a random-walk. 

	All the considerations above lead to these conclusions:
	\begin{itemize}
	\item
		The \thealg{} routing algorithm is fast and efficient. Routing queries
		in the  direction of the semantic group that can satisfy them is
		a winning strategy. 
	\item
		The \thealg{} links management algorithm allows similar peers to
		form  high connected clusters. This fact allows
		queries to be answered faster and more efficiently \footnote{In
		terms of number of walks needed to satisfy a query}
	\end{itemize}
	
	\section{Future works}
	\label{sect:future}
	
	In this paper a novel P2P self--organising algorithm for  resource 
	searching and retrieving  has been presented. The algorithm
  emulates the way social relationships among people naturally
  arise and evolve, and finally produces a really small--world network
	topology, as confirmed by simulation results.
	\thealg{} is a valid alternative to actual P2P 	structures
	based on  simple flooding or random walk. In fact	similar peers in
	\thealg{} form strong interconnected
	``semantic--groups'', allowing fast and efficient query routing.
	Future work will focus on extending \thealg{} in order to include
	other mechanisms 	typically found in social communities, such as:
	\begin{itemize}
	\item
		Random meetings among peers (that allows peer to connect, via ALs,
		also to	non--similar peers)
	\item
		Advertising of new resources
	\item
		Semantic--Links scoring, to simulate various possible  degrees of
		acquaintance among people
	\end{itemize}

	\bibliography{article}
	\bibliographystyle{plain}
	

\end{document}